\documentclass[twoside,twocolumn,9pt]{article} 
\usepackage{extsizes}
\usepackage[super,sort&compress,comma]{natbib}
\usepackage[version=3]{mhchem}
\usepackage[left=1.5cm, right=1.5cm, top=1.785cm, bottom=2.0cm]{geometry}
\usepackage{balance}
\usepackage{mathptmx}
\usepackage{sectsty}
\usepackage{graphicx}
\usepackage{lastpage}
\usepackage[format=plain,justification=justified,singlelinecheck=false,font={stretch=1.125,small,sf},labelfont=bf,labelsep=space]{caption}
\usepackage{float}
\usepackage{fancyhdr}
\usepackage{fnpos}
\usepackage[english]{babel}
\addto{\captionsenglish}{%
  
}
\usepackage{array}
\usepackage{charter}
\usepackage[T1]{fontenc}
\usepackage[usenames,dvipsnames]{xcolor}
\usepackage{setspace}
\usepackage[compact]{titlesec}
\usepackage{hyperref}
\usepackage{epstopdf}

\definecolor{cream}{RGB}{222,217,201}

\usepackage{amsmath,amssymb} 

\begin{document}

\pagestyle{fancy}
\thispagestyle{plain}
\fancypagestyle{plain}{
\renewcommand{\headrulewidth}{0pt}
}

\makeFNbottom
\makeatletter
\renewcommand\LARGE{\@setfontsize\LARGE{15pt}{17}}
\renewcommand\Large{\@setfontsize\Large{12pt}{14}}
\renewcommand\large{\@setfontsize\large{10pt}{12}}
\renewcommand\footnotesize{\@setfontsize\footnotesize{7pt}{10}}
\makeatother

\renewcommand{\thefootnote}{\fnsymbol{footnote}}
\renewcommand\footnoterule{\vspace*{1pt}%
\color{cream}\hrule width 3.5in height 0.4pt \color{black}\vspace*{5pt}}
\setcounter{secnumdepth}{5}
\makeatletter
\renewcommand\@biblabel[1]{#1}
\renewcommand\@makefntext[1]%
{\noindent\makebox[0pt][r]{\@thefnmark\,}#1}
\makeatother
\renewcommand{\figurename}{\small{Fig.}~}
\sectionfont{\sffamily\Large}
\subsectionfont{\normalsize}
\subsubsectionfont{\bf}
\setstretch{1.125} 
\setlength{\skip\footins}{0.8cm}
\setlength{\footnotesep}{0.25cm}
\setlength{\jot}{10pt}
\titlespacing*{\section}{0pt}{4pt}{4pt}
\titlespacing*{\subsection}{0pt}{15pt}{1pt}

\fancyfoot{}
\fancyfoot[LO,RE]{\vspace{-7.1pt}\includegraphics[height=9pt]{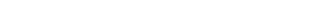}}
\fancyfoot[CO]{\vspace{-7.1pt}\hspace{13.2cm}\includegraphics{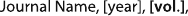}}
\fancyfoot[CE]{\vspace{-7.2pt}\hspace{-14.2cm}\includegraphics{head_foot/RF}}
\fancyfoot[RO]{\footnotesize{\sffamily{1--\pageref{LastPage} ~\textbar  \hspace{2pt}\thepage}}}
\fancyfoot[LE]{\footnotesize{\sffamily{\thepage~\textbar\hspace{3.45cm} 1--\pageref{LastPage}}}}
\fancyhead{}
\renewcommand{\headrulewidth}{0pt}
\renewcommand{\footrulewidth}{0pt}
\setlength{\arrayrulewidth}{1pt}
\setlength{\columnsep}{6.5mm}
\setlength\bibsep{1pt}

\makeatletter
\newlength{\figrulesep}
\setlength{\figrulesep}{0.5\textfloatsep}

\newcommand{\topfigrule}{\vspace*{-1pt}%
\noindent{\color{cream}\rule[-\figrulesep]{\columnwidth}{1.5pt}} }

\newcommand{\botfigrule}{\vspace*{-2pt}%
\noindent{\color{cream}\rule[\figrulesep]{\columnwidth}{1.5pt}} }

\newcommand{\dblfigrule}{\vspace*{-1pt}%
\noindent{\color{cream}\rule[-\figrulesep]{\textwidth}{1.5pt}} }

\makeatother

\twocolumn[
  \begin{@twocolumnfalse}
{\includegraphics[height=30pt]{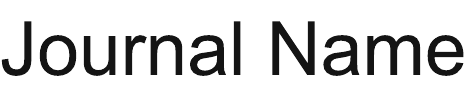}\hfill\raisebox{0pt}[0pt][0pt]{\includegraphics[height=55pt]{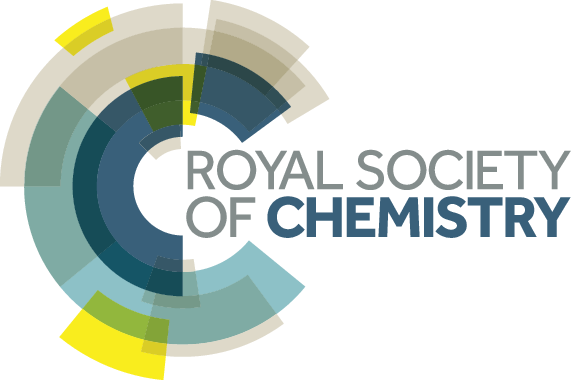}}\\[1ex]
\includegraphics[width=18.5cm]{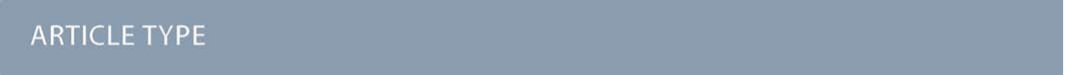}}\par
\vspace{1em}
\sffamily
\begin{tabular}{m{4.5cm} p{13.5cm} }

\includegraphics{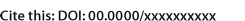} & \noindent\LARGE{\textbf{Increasing the Earth's Albedo: The K\"{o}hler Equation at Sea}} \\
\vspace{0.3cm} & \vspace{0.3cm} \\

& \noindent\large{J. I. Katz$^{\ast}$\textit{$^{a}$}} \\

\includegraphics{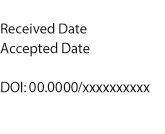} & \noindent\normalsize{
Increasing marine haze and clouds has been considered as a possible means of
increasing the Earth's albedo.  This would reduce Solar heating and global
warming, counteracting the effects of the anthropogenic increase in
greenhouse gases.  One proposed method of doing so would inject small
droplets of seawater or condensation nuclei into the marine boundary layer,
creating artificial haze and cloud.  The equilibrium size of such droplets
is described by the K\"{o}hler equation that includes the vapor pressure
reduction attributable to the solute according to Raoult's law and the vapor
pressure increase of a small droplet as a result of surface tension
according to Kelvin.  Here we apply this classic result to small droplets
in the marine boundary layer, where the partial pressure of water vapor is
less than the equilibrium vapor pressure because it is in equilibrium with
the saline ocean.  We calculate the equilibrium size of a droplet containing
dissolved ions and find that the radius of a droplet of seawater shrinks
greatly before it achieves equilibrium.} \\

\end{tabular}

 \end{@twocolumnfalse} \vspace{0.6cm}

  ]

\renewcommand*\rmdefault{bch}\normalfont\upshape
\rmfamily
\section*{}
\vspace{-1cm}


\footnotetext{\textit{$^{a}$~Department of Physics and McDonnell Center for the Space Sciences, Washington University, St. Louis, Mo. USA. Fax: 314-935-6219; Tel: 314-935-6202; E-mail: katz@wuphys.wustl.edu}}
\section{Introduction}
Marine cloud-brightening by injection of seawater aerosols as condensation
nuclei was proposed \cite{L90,L02,L12} to increase the Earth's albedo
(fraction of incident sunlight reflected back to space) to counteract the
warming effects of anthropogenic greenhouse gases.  Here I review the
properties of saline water droplets in equilibrium with the oceanic boundary
layer that is itself in equilibrium with the ocean.  K\"{o}hler's classic
equation describes the vapor pressure of a small droplet, increased by
surface tension and decreased by the presence of a solute.  The equilibrium
size of such a droplet depends on the ambient vapor density.  Here I apply
those results to droplets, initially with the salinity of seawater, in
equilibrium with the water vapor in the marine boundary layer.  This water
vapor is less dense than saturated water vapor because it is in equilibrium
with the saline ocean, whose vapor pressure is reduced by its dissolved
salts.
\section{K\"{o}hler's Equation}
K\"{o}hler's equation for the vapor pressure $p(r)$ of a droplet of radius
$r$ of fluid with molecular number density $n$ and surface tension $\sigma$,
containing $N$ solute atoms, ions or molecules \cite{RY89,Y93,WH77,A10} may
be written
\begin{equation}
\label{kohler}
\ln{\left({p(r) \over p_0(\infty)}\right)} = {2 \sigma \over k_B T n r}
- {3 \over 4 \pi}{N \over n r^3},
\end{equation}
where $p_0(\infty)$ is the vapor pressure of a flat surface of the pure
liquid.  This result is based on Raoult's law \cite{LL58} for the
equilibrium partial pressure $p_i$ of a component $i$ that has equilibrium
vapor pressure $p_0$ in a solution in which $i$ constitutes a mole fraction
$x_i$
\begin{equation}
	\label{raoult}
	p_i = p_0 x_i
\end{equation}
and Kelvin's law \cite{P66}
\begin{equation} 
	\label{kelvin}
	p(r) = p_0(\infty) - \sigma \left({\rho_{vapor} \over
	\rho_{liquid}-\rho_{vapor}}\right) {2 \over r}
\end{equation}
for the reduction of vapor pressure by curvature of the surface of a
spherical drop of radius $r$ and that these small corrections are additive.
It neglects non-ideal properties of the solution and any dependence of
$\sigma$ on the solute (all good approximations for dilute solutions).  $N$
must include the van't Hoff factor, the number of dissociated ions per
molecule, which is close to 2 for dilute solutions of ocean salts.
\section{K\"{o}hler's Equation at Sea}
For droplets in the atmospheric sea-surface boundary layer in equilibrium
with slightly undersaturated water vapor, the application of K\"{o}hler's
equation requires consideration of the undersaturation.  Substituting $p(r)
= p(\infty) = (1 - \epsilon) p_0(\infty)$ in the left hand side of
Eq.~\ref{kohler} and approximating $\ln{(1-\epsilon)} \approx - \epsilon$
yields a cubic equation for the equilibrium radius $r$:
\begin{equation}
\label{kohleratsea}
	r^3 + {2 \sigma \over k_B T n \epsilon}r^2 - {3 \over 4 \pi}{N \over
	\epsilon n} = 0.
\end{equation}
We take $T = 293$ K, $n = 3.35 \times 10^{22}$ cm$^{-3}$ and $\sigma = 70$
ergs/cm$^2$.

The atmospheric boundary layer over the ocean is in equilibrium with saline
seawater \cite{sea}.  Application of Raoult's law (Eq.~\ref{raoult}) to the
vapor pressure $p(\infty)$ of a flat surface of seawater indicates that
$p(\infty) = 0.9796 p_0(\infty)$; we define the undersaturation $\epsilon =
1 - p(\infty)/p_0(\infty) = 1 - 0.9796 = 0.0204$.
\section{Nucleation}
\subsection{Minimum $N$ and $r$}
The first two terms in Eq.~\ref{kohleratsea} are positive while the third
term, $\propto N$, is negative.  This sets a lower bound on the values of
$N$ for which a solution is possible.  This bound is not zero because
the minimum meaningful $r_{min} \approx 10^{-8}$ cm (for a smaller droplet
$\sigma$ is not well defined).  For small $r$ the cubic term may be
neglected in comparison to the quadratic term, yielding
\begin{equation}
	\label{Nmin}
	N \ge {8 \pi \over 3} {\sigma \over k_B T} r_{min}^2 \gtrsim 1.5.
\end{equation}

Although the formalism indicates that for some values of the parameters
there is a meaningful minimum $N$ that permits formation of a stable
droplet, for droplets large enough to justify the use of continuum theory
this minimum (Eq.~\ref{Nmin}) is so small as to be irrelevant.  It is
meaningful to solve Eq.~\ref{kohleratsea} for $r(N)$ or $N(r)$ when $r \gg
10^{-8}$ cm and $N \gg 2$.
\subsection{Equilibrium Droplet Size}
\begin{figure}
\centering
\includegraphics[width=3.5in]{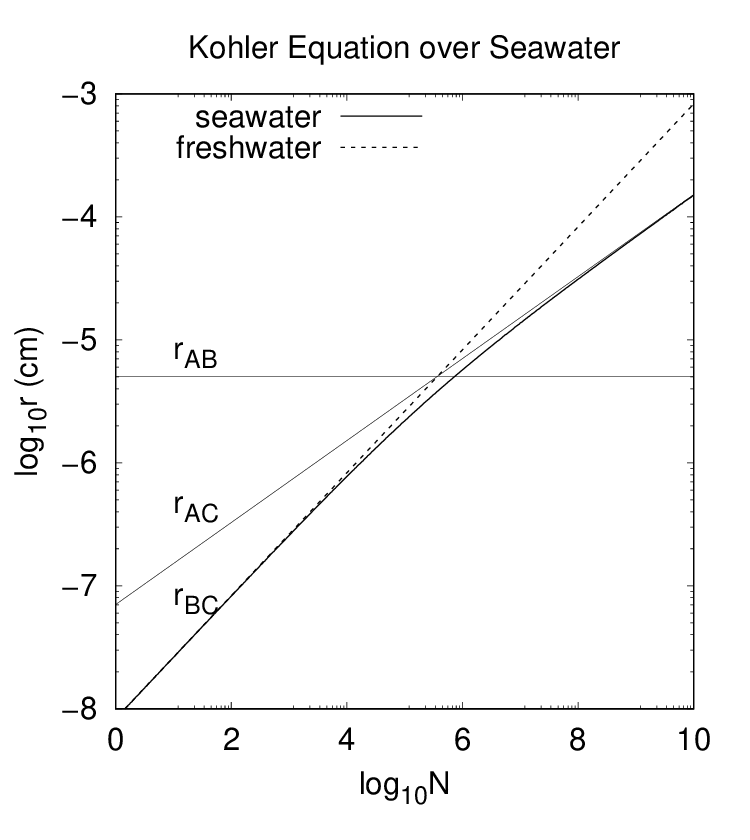}
\caption{\label{kohlerfig} Relation between equilibrium droplet radius $r$
and nunber of solute ions $N$, evaluated in the marine atmospheric boundary
layer.  The air is slightly undersaturated in water vapor compared to air
over pure water because of ocean salinity.  At smaller $N$ and $r$ the
quadratic (surface tension) term in Eq.~\ref{kohleratsea} dominates and $r
\propto N^{1/2}$, while at larger $N$ and $r$ the cubic (undersaturation)
term dominates and $r \propto N^{1/3}$.  The cross-over between these
regimes occurs for $r \approx r_{AB} \approx 5 \times 10^{-6}\,$cm and $N
\approx 5 \times 10^5$ (Eq.~\ref{AB}).  $r_{BC}$ is identical to $r(N)$ for
fresh water.}
\end{figure}

The cubic Eq.~\ref{kohleratsea} may be solved explicitly for $r(N)$, giving
the equilibrium radius of a drop nucleated by a particle containing $N$
nonvolatile solute atoms, ions or molecules.  The cumbersome algebraic
solution is shown graphically in Fig.~\ref{kohlerfig}.  A droplet with
radius $r < r(N)$ will condense additonal water vapor until $r = r(N)$.  A
droplet with $r > r(N)$ will evaporate until $r = r(N)$.

There are three cross-overs in Eq.~\ref{kohleratsea}, values of $r$ when two
terms are equal.  The cubic term (in $r$) equals the quadratic term when
\begin{equation}
	\label{AB}
	r = r_{AB} = {2 \sigma \over k_B T n \epsilon} \approx 5.06 \times
	10^{-6}\ \text{cm}.
\end{equation}
Each of these terms act to make the droplet evaporate, but for $r > r_{AB}$
subsaturation is more important that surface tension.  If there is
sufficient solute these effects may be overcome and a droplet condenses
further water vapor.  To simplify the algebra we consider these terms
separately.

The cubic term in $r$ in Eq.~\ref{kohleratsea} equals the negative term when
\begin{equation}
	\label{AC}
	r = r_{AC} = \left({3 \over 4\pi}{N \over n\epsilon}\right)^{1/3}
	\approx 7.04 \times 10^{-8} N^{1/3}\ \text{cm}.
\end{equation}
When $r > r_{AC}$ subsaturation overcomes the effect of the solute and
the droplet shrinks.  If $r < r_{AB}$ surface tension also contributes to
evaporation.  If $r < r_{AC}$ the solute is more important and the droplet
will condense additional water vapor unless $r < r_{AB}$, in which case
surface tension must also be considered.

The term quadratic in $r$ equals the negative term when
\begin{equation}
	\label{BC}
	r = r_{BC} = \sqrt{{3 \over 8\pi}{N k_B T \over \sigma}} \approx
	8.30 \times 10^{-9} N^{1/2}\ \text{cm}.
\end{equation}
When $r > r_{BC}$ the droplet will shrink, while for $r < r_{BC}$ the
droplet will condense additional water vapor, unless $r > r_{AC}$.   If
$r > r_{AC}$ subsaturation makes the droplet shrink, even without the
contributing effect of surface tension.

Very small droplets (those with $r \ll 10^{-5}\,$cm ) are ineffective
scatterers of light, so we now consider the regime in which the quadratic
term in Eq.~\ref{kohleratsea} may be neglected in comparison to the cubic
term ($r \gg r_{AB} \approx 5 \times 10^{-6}\,$cm).  Neglecting the
quadratic term, we find
\begin{equation}
	r^3 \approx {3 \over 4 \pi}{N \over n\epsilon}.
\end{equation}
Then the number of water molecules in the droplet
\begin{equation}
	N_{\text{H}_2\text{O}} \approx {N \over \epsilon} \approx 50 N.
\end{equation}
Condensation of seawater multiplies the volume of the aerosol by a factor
${\cal O}(50)$ and multiplies its scattering by another large factor.  This
latter factor is proportional to the square of the volume in the Rayleigh
scattering regime applicable to very small droplets ($r \lesssim
10^{-5}\,$cm), and to the 2/3 power of the volume in the geometrical optics
regime applicable to larger droplets ($r \gtrsim 10^{-4}\,$cm) \cite{H81}).
\section{Seawater Droplets}
Rather than injecting small soluble particles to nucleate droplets from
water vapor in the marine boundary layer, it is possible to inject small
droplets of seawater, available in unlimited quantity at sea.  Such a
particle will have a higher vapor pressure than the sea itself because of
its surface curvature.  Equate the vapor pressure in a droplet of seawater
that has shrunken by evaporation from an initial radius $r_0$ (containing
$N = 4 \pi n \epsilon r_0^3/3$ solute ions) to a radius $r$ to the partial
pressure of water over the ocean:
\begin{equation}
	p_\infty\left(1 - \epsilon {r_0^3 \over r^3}\right) +
	{2 \sigma \over r} = p_\infty (1-\epsilon),
\end{equation}
where $p_\infty \approx 1.7 \times 10^4$ dyne/cm$^2$ is the vapor pressure
of a flat surface of pure water at 15$^\circ$C.  Defining a characteristic
radius
\begin{equation}
	R_0 = {2 \sigma \over p_\infty \epsilon} \approx 0.4\ \text{cm}
\end{equation}
leads to the cubic equation
\begin{equation}
	\label{kohleratsea2}
	r^3 + R_0 r^2 - r_0^3 = 0.
\end{equation}

In the relevant regime $r \ll R_0$ (larger droplets fall rapidly as rain)
and the cubic term may be neglected, leading to the approximation
\begin{equation}
	\label{shrinkage}
	{r \over r_0} \approx \sqrt{r_0 \over R_0} \ll 1.
\end{equation}
The increase in vapor pressure attributable to surface tension is balanced
for the shrinking droplet by the decrease resulting from increasing
salinity.  The Raoult (solute) decrease grows faster with decreasing radius
than does the Kelvin (surface tension) increase, and the droplet comes to an
equilibrium radius that is much smaller than its initial radius.

\begin{figure}
	\centering
	\includegraphics[width=3.4in]{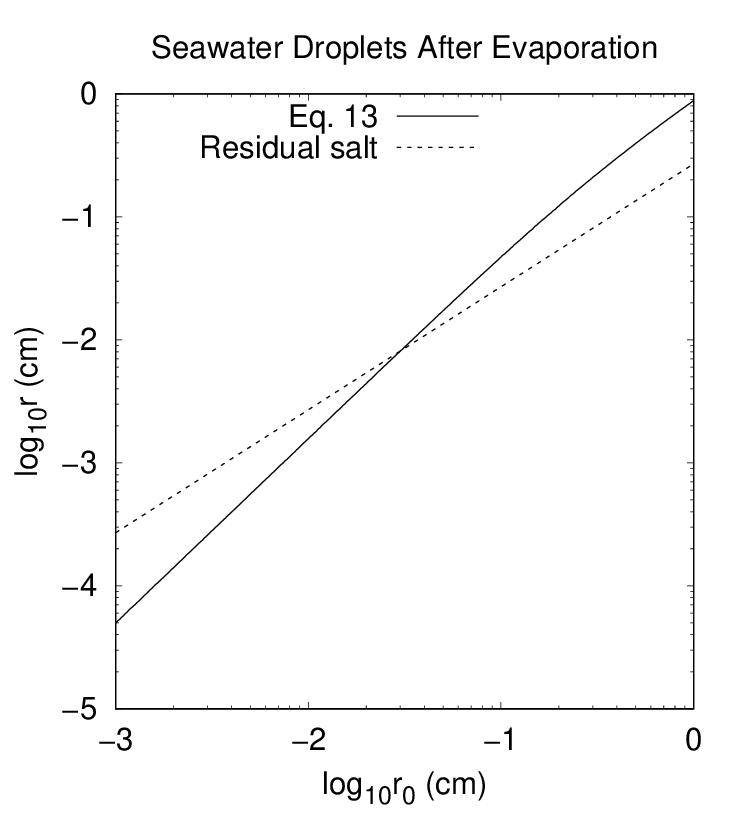}
	\caption{\label{kohlerf2} Equilibrium radius $r$ of seawater drop
	of original radius $r_0$ in oceanic boundary layer according to
	Eq.~\ref{kohleratsea2}.  Dashed line ($r = 0.27 r_0$) shows where
	Raoult's law breaks down becauze the droplet becomes concentrated
	brine or moist salt; values of $r$ lying below this line are
	unphysical.}
\end{figure}

The solution of Eq.~\ref{kohleratsea2} is shown in Fig.~\ref{kohlerf2}.
In fact, shrinkage is limited by the breakdown of Raoult's law when the
solution becomes concentrated.  $r/r_0$ cannot be less than ${\cal O}
(\epsilon^{1/3}) \approx 0.27$; the result is a particle of moist salt or
highly concentrated brine containing all the originally dissolved salt but
little or none of the original water.  Its hygroscopicity is more than
offset by the increase of its vapor pressure by surface tension.

\section{Conclusions}
Increasing the Earth's albedo by stimulating cloud formation has been
proposed as a method of counteracting the global warming produced by
anthropogenic greenhouse gases.  Condensation of water vapor on a salt
nucleus multiplies its mass, volume and optical scattering by large factors
(the mechanism by which cloud seeding may stimulate rainfall from clouds
supersaturated by evaporation from small droplets).  Although the ready
availability of seawater may make its atomization appear to be an attractive
means of producing scattering clouds, the increased vapor pressure of small
seawater droplets makes them shrink by large factors in radius (and by the
cube of that factor in mass and volume) as they come into equilibrium with
the ambient water vapor in the marine boundary layer.
\section*{Author Contributions}
The single author conceived and carried out the research and wrote the
paper. 
\section*{Conflicts of Interest}
There are no conflicts of interest.
\section*{Acknowledgments}
I thank Matthew Alford for useful discussions.
\bibliography{kohlerEES}
\bibliographystyle{rsc}
\end{document}